\documentclass{ws-procs9x6}
\newcommand{\be}{\begin{equation}}
\newcommand{\ee}{\end{equation}}
\newcommand{\bea}{\begin{eqnarray}}
\newcommand{\eea}{\end{eqnarray}}
\newcommand{\bean}{\begin{eqnarray*}}
\newcommand{\eean}{\end{eqnarray*}}
\newcommand{\ba}{\begin{array}}
\newcommand{\ea}{\end{array}}
\newcommand{\non}{\nonumber}
\newcommand{\bc}{\begin{center}}
\newcommand{\ec}{\end{center}}

\begin{document}

\title{Transverse double-spin asymmetries \\
for small $Q_T$ Drell-Yan pair production \\
in $pp$ and $p\bar{p}$ collisions}

\author{H.~Kawamura}
\address{Radiation Laboratory, RIKEN,
Wako 351-0198, Japan\\
E-mail: hiroyuki@rarfaxp.riken.jp}

\author{J.~Kodaira}
\address{Theory Division, KEK,
Taukuba 305-0801, Japan\\
E-mail: jiro.kodaira@kek.jp}

\author{K.~Tanaka}
\address{Department of Physics, Juntendo University,
Inba, Chiba 270-1695, Japan\\
E-mail: tanakak@sakura.juntendo.ac.jp}  

\maketitle
\abstracts{ 
We discuss transverse double-spin asymmetries
for dimuon production at small transverse-momentum $Q_T$ in $pp$ and 
$p\bar{p}$ collisions. 
All order resummation of large logarithms 
relevant in small $Q_T$ region is performed at 
next-to-leading logarithmic (NLL) accuracy, and asymmetries 
at RHIC, J-PARC and GSI are calculated.}
\vspace{-0.5cm}

The transversity $\delta q(x)$, the distribution of transversely
polarized quarks inside transversely polarized nucleon, is the last 
unknown distribution of nucleon at the leading twist.\cite{RS:79}  
It is not measurable
in inclusive DIS due to its chiral-odd structure, 
and a number of experiments are underway to measure it through
semi-inclusive processes. 
Transversely polarized Drell-Yan process (tDY) is another way 
to measure it, which could in principle provide us with clean
information on the transversity.
However, the actual feasibility of extracting $\delta q(x)$ 
from tDY data depends on the magnitude of transverse double-spin 
asymmetry: 
$\frac{\Delta_Td\sigma}{d\sigma}\equiv
\frac{d\sigma^{\uparrow\uparrow}-d\sigma^{\uparrow\downarrow}}
{d\sigma^{\uparrow\uparrow}+d\sigma^{\uparrow\downarrow}}$.
At RHIC, the asymmetry is possibly small since (i) $pp$ collision 
probes products of valence-quark and sea-antiquark distributions where 
the latter are supposed to be small, and (ii) the small-$x$ rise of 
unpolarized sea-quark distributions enhance the denominator of 
the asymmetry at high energy.\cite{MSSV:98}
On the other hand, possibilities of transversly polarized 
$pp$ ($p\bar{p}$) experiments at J-PARC (GSI) with moderate energies 
are recently discussed, 
where larger asymmetries in tDY are expected.\cite{Shimizu:2005fp}
In this paper, we explore the asymmetries in $Q_T$ spectrum of Drell-Yan 
pair, especially in small $Q_T$ region where the bulk of the lepton pair 
is produced.

In the small $Q_T$ region, fixed-order perturbative calculation 
does not work well 
since there appear large logarithmic corrections at each order of 
perturbation theory as $\alpha_s^n\ln^{2n-1}(Q^2/Q_T^2)/Q_T^2$, 
$\alpha_s^n\ln^{2n-2}(Q^2/Q_T^2)/Q_T^2$, and so on, 
corresponding to LL, NLL, and higher contributions, respectively.
These ``recoil logs'' have to be resummed to all orders in $\alpha_s$ 
to obtain reliable perturbative predictions. 
The resummation is carried out in the impact parameter $b$ space,
conjugate to the $Q_T$ space, to take transverse-momentum conservation 
into account, and the resummed cross section is expressed as 
the Fourier transform back to the $Q_T$ space.\cite{CSS:85} 
At the NLL, the resummed cross section of tDY, 
differential in invariant mass $Q$, transverse momentum $Q_T$ 
and rapidity $y$ of the lepton pair, and in
azimuthal angle $\phi$ of one of the outgoing leptons is given 
by\cite{KKST:05} ($\sqrt{S}$ the CM energy of hadrons)
\bea
\frac{\Delta_T d \sigma^{\rm NLL}}{d Q^2 d Q_T^2 d y d \phi}
=&&\!\!
\cos(2 \phi )
\frac{\alpha^2}{3\, N_c\, S\, Q^2}
\sum_{i}e_i^2 \int_0^{\infty} d b \frac{b}{2}
J_0 (b Q_T)e^{\, S (b , Q)}
\label{resum}
\\
\times&&\!\!\!
\Biggl[  ( C_{qq} \otimes \delta q_i )
           \left( x_1^0 , \frac{b_0^2}{b^2} \right)      
( C_{\bar{q} \bar{q}} \otimes \delta\bar{q}_i )
           \left( x_2^0 , \frac{b_0^2}{b^2} \right)
+ ( x_1^0 \leftrightarrow x_2^0 )\Biggr].
\non
\eea 
Here  
$J_0 (b Q_T)$ is a Bessel function, 
$b_0 = 2e^{-\gamma_E}$ with $\gamma_E$ the Euler constant, and
the large logarithmic corrections are resummed into 
the Sudakov factor $e^{S (b , Q)}$ with 
$S(b,Q)=-\int^{Q^2}_{b_0^2/b^2}(d\kappa^2/\kappa^2)
\{\ln\frac{Q^2}{\kappa^2}A_q(\alpha_s(\kappa))+B_q(\alpha_s(\kappa))\}$.
$A_q$, $B_q$ and the coefficient functions $C_{qq},C_{\bar{q}\bar{q}}$
are perturbatively calculable, and are found in Ref.\cite{KKST:05} 
up to the accuracy necessary for the NLL.
$x_{1,2}^0 = (Q/\sqrt{S})e^{\pm y}$, 
and $\delta q_i(x,\mu^2)$ is the transversity of $i$-th flavour quark 
at the $\overline{\rm MS}$ scale $\mu$. 
The singularity in $b$-integration,\cite{CSS:85} due to 
the Landau pole in $\alpha_s(\kappa)$,
is taken care of by ``contour deformation method'' introduced in 
the joint resummation;\cite{LKSV:01} correspondingly,
\cite{CSS:85,BCDeG:03,LKSV:01} 
we take non-perturbative effects into account by the 
replacement $e^{S (b , Q)} \rightarrow e^{S (b , Q)-g_{NP}b^2}$
in (\ref{resum}), with a non-perturbative parameter $g_{NP}$.
We combine the resulting NLL cross section with the leading order (LO) 
cross section, which is of $\mathcal{O}(\alpha_s )$ and is 
obtained\cite{KKST:05} as QCD prediction at large $Q_T$;
the matching of the NLL formula with the corresponding component 
in the LO cross section is performed at intermediate $Q_T$
following the formulation developed in Ref.\cite{BCDeG:03}, to ensure 
no double counting for all $Q_T$,
and we finally obtain the complete ``NLL+LO'' cross section 
$\Delta_T d \sigma /(d Q^2 d Q_T^2 d y d \phi )$,
which has a uniform accuracy over the entire range of 
$Q_T$.\cite{KKST:05}  
Note that, 
also for the unpolarized DY, the corresponding ``NLL+LO'' cross section 
$d \sigma /(d Q^2 d Q_T^2 d y d \phi )$ can be obtained 
in the same framework,
utilizing the results in the literatures.\cite{AEGM:84,CSS:85}.

We calculate the following transverse double-spin asymmetries: 
\bea
A_{TT}=\left[\Delta_T d \sigma / d Q^2 d Q_T^2 d y d \phi\right]
\left/
\left[ d \sigma / d Q^2 d Q_T^2 d y d \phi \right]\ .
\right.
\label{asym}
\eea
As non-perturbative inputs, we use the same parton distributions 
as those used in Ref.\cite{MSSV:98};
in particular, for the numerator,
we use a model of the transversity $\delta q(x,\mu^2)$ 
which saturates the Soffer bound as
$\delta q(x,\mu_0^2)=[q(x,\mu_0^2)+\Delta q(x,\mu_0^2)]/2$
at low input scale $\mu_0\sim 0.6$ GeV and is evolved to higher $\mu^2$ 
with NLO DGLAP kernel.\cite{KMHKKV:97}
The non-perturbative parameter $g_{NP}$ is taken to be the same value 
for both numerator and denominator of (\ref{asym}), 
and we use $g_{NP} \simeq 0.5$ GeV$^2$ 
as suggested by the result of Ref.\cite{KS}. 
\begin{figure}[ht]
\centerline{\hspace{-0.2cm}
\epsfxsize=5cm   
\epsfbox{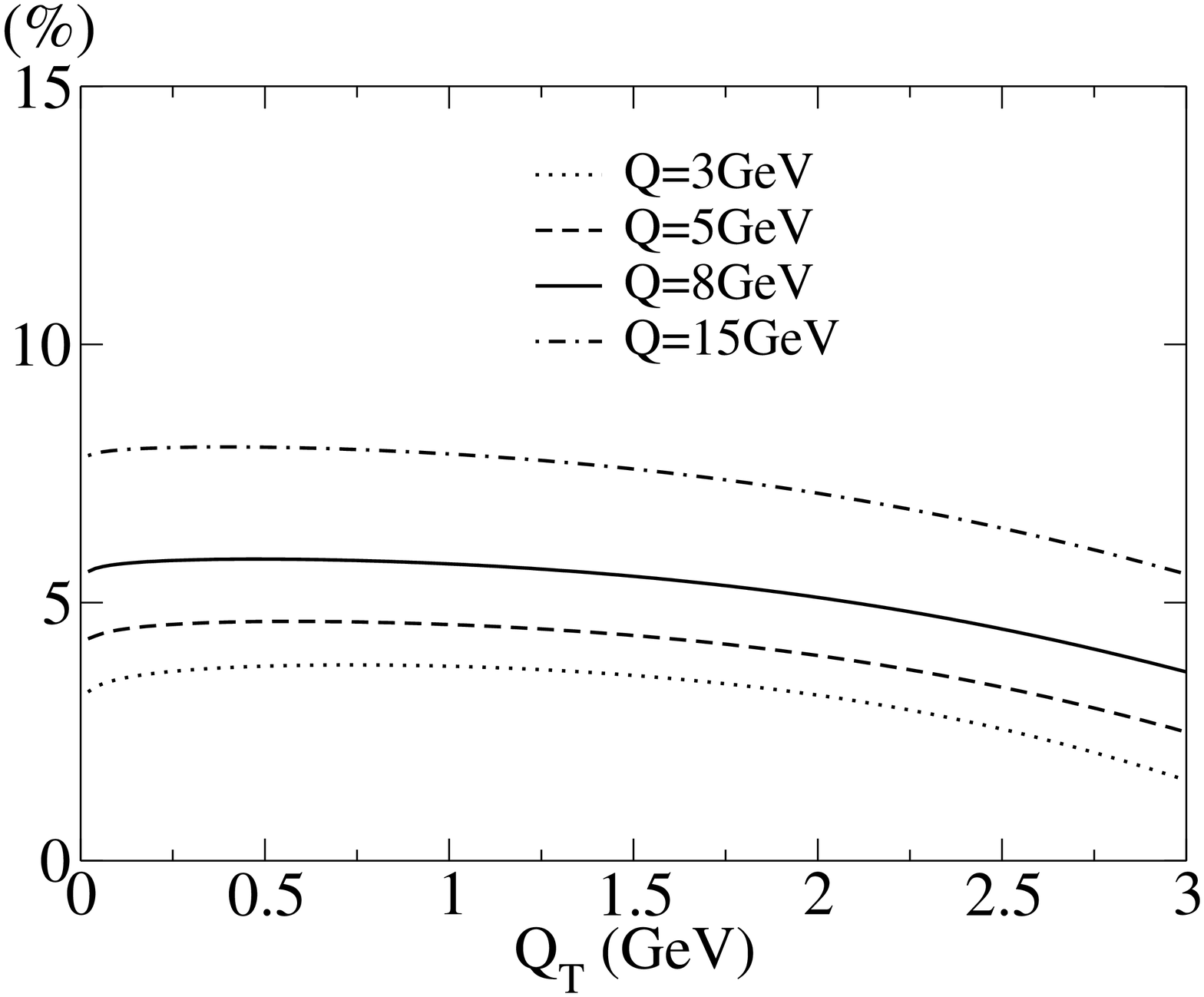}
\hspace{0.4cm}
\epsfxsize=5cm
\epsfbox{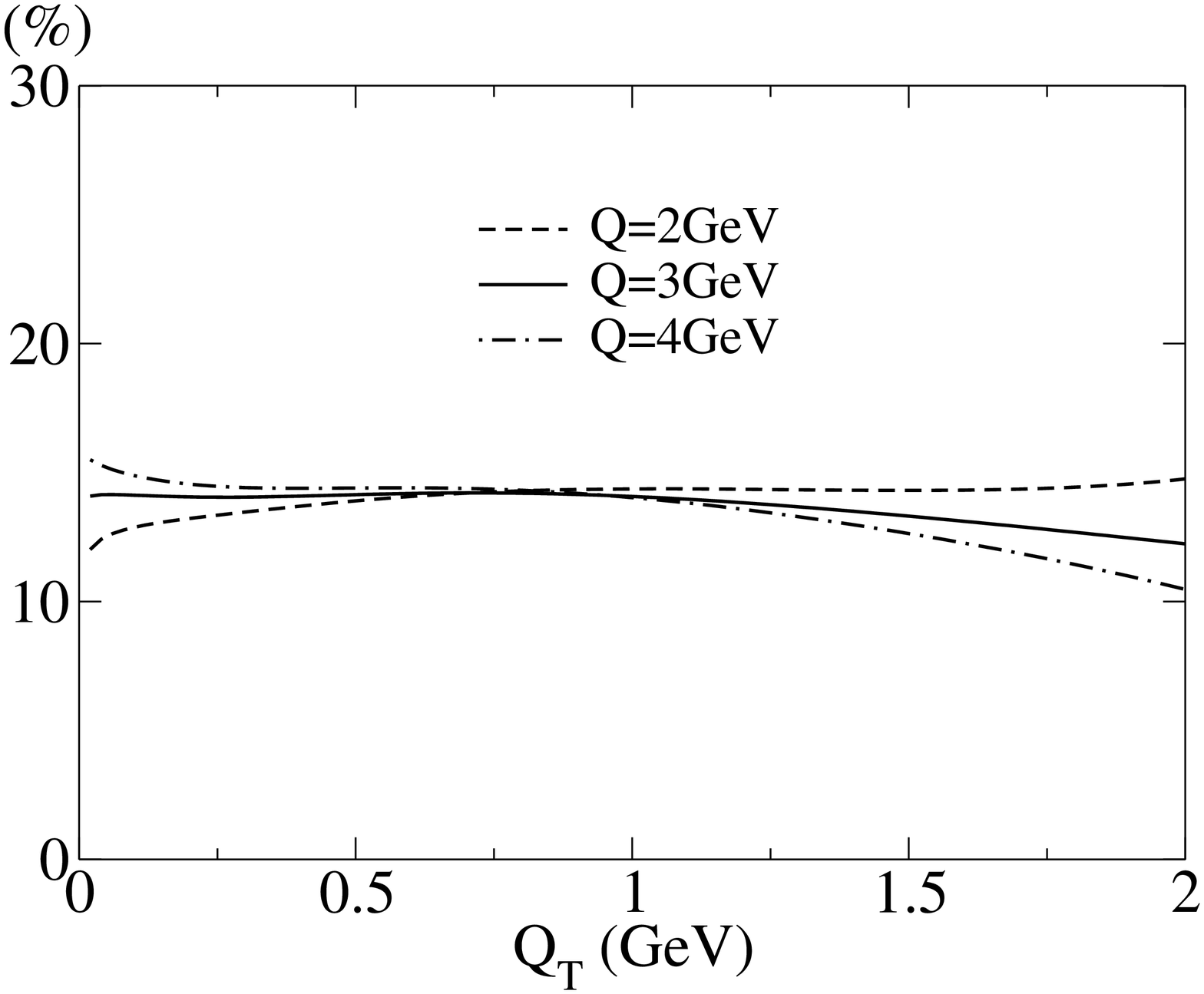}}
\caption{$A_{TT}$ in $pp$ collision.
Left: $\sqrt{S}=200$ GeV, $y=2$, $\phi=0$ and 
$Q=3,5,8,15$ GeV, corresponding to RHIC kinematics. 
Right: $\sqrt{S}=10$ GeV, $y=\phi=0$ and $Q=2,3,4$ GeV, 
corresponding to J-PARC kinematics. $g_{NP}=0.5$ GeV$^2$ in both cases.} 
\end{figure}
\vspace{-0.2cm}

$A_{TT}$ in $pp$ collision are shown as functions of $Q_T$ in Fig.~1, 
where the left (right) panel is for RHIC (J-PARC) kinematics.
$A_{TT} \gtrsim 10$\% are obtained for J-PARC, where the parton 
distributions are probed at medium $x_{1,2}^{0}$ (see (\ref{resum})). 
For the RHIC case, $A_{TT}$ are less than 10\%, 
and becomes smaller for smaller $Q$ due to the small-$x$ rise of 
unpolarized sea-distributions in the denominator of (\ref{asym}).
\begin{figure}[ht]
\centerline{\epsfxsize=5.5cm,\epsfbox{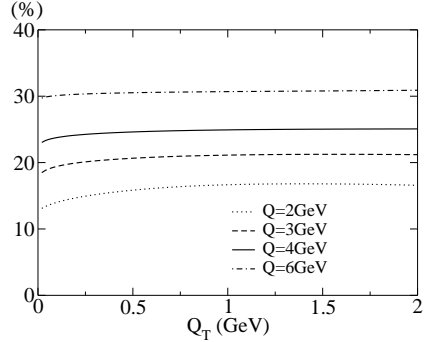}}
\caption{$A_{TT}$ in $p\bar{p}$ collision at
$\sqrt{S}=14.5$ GeV, $y=\phi=0$,
and $Q=2,3,4,6$ GeV, corresponding to GSI kinematics, and with 
$g_{NP}=0.5$GeV$^2$.} 
\vspace{-0.2cm}
\end{figure}
The largest $A_{TT}$ of 15-30\% are obtained in $p\bar{p}$ collision 
at GSI kinematics 
as shown in Fig.~2, where $A_{TT}$ are dominated by valence
distributions at medium $x$.
Integrating the numerator and denominator 
of (\ref{asym}) over $Q_T^2$, our results for GSI kinematics
reproduce the corresponding NLO asymmetries given by Barone et
al.\cite{Shimizu:2005fp}. 

In all cases of Figs.~1 and 2, $A_{TT}$ have flat behavior
in small $Q_T$ region, 
although the numerator and denominator of (\ref{asym}) have strong $Q_T$ 
dependence, respectively.\cite{KKST:05}
We also note that $A_{TT}$ are almost unaffected
when the parameter $g_{NP}$ is varied in the range 
$g_{NP}=0.3$-0.8 GeV$^2$. 
These common features in Figs.~1 and 2 come from the dominance of 
soft gluon effects in small $Q_T$ region, whose main part, 
i.e. the Sudakov factor of (\ref{resum}), is universal\cite{KKST:05} 
at the NLL level in both polarized and unpolarized channels.

\vspace{-0.1cm}
\section*{Acknowledgments}
We would like to thank Werner Vogelsang and Stefano Catani   
for valuable discussions.
The work of J.K. and K.T. was supported by the Grant-in-Aid for
Scientific Research Nos. C-16540255 and C-16540266.

\vspace{-0.1cm}


\end{document}